\documentclass[aps,twocolumn,pra,showpacs,floatfix]{revtex4}

\usepackage{amssymb,amsmath}
\usepackage{graphics,epsfig}

\newcommand{\eqnref}[1]{Eq.~(\ref{#1})}
\newcommand{\eqnsref}[1]{Eqs.~(\ref{#1})}
\newcommand{\Eqnref}[1]{Equation~(\ref{#1})}

\newcommand{\tblref}[1]{Table~\ref{#1}}

\newcommand{\figref}[1]{Fig.~\ref{#1}}

\newcommand{\secref}[1]{Sec.~\ref{#1}}
\newcommand{\Secref}[1]{Section~\ref{#1}}

\newcommand {\bfa}  {{\mathbf{a}}}
\newcommand {\bfk}  {{\mathbf{k}}}
\newcommand {\bfl}  {{\mathbf{l}}}
\newcommand {\bfp}  {{\mathbf{p}}}
\newcommand {\bfr}  {{\mathbf{r}}}
\newcommand {\bfs}  {{\mathbf{s}}}
\newcommand {\bfu}  {{\mathbf{u}}}
\newcommand {\bfv}  {{\mathbf{v}}}
\newcommand {\bfx}  {{\mathbf{x}}}
\newcommand {\bfz}  {{\mathbf{z}}}
\newcommand {\bfA}  {{\mathbf{A}}}
\newcommand {\bfB}  {{\mathbf{B}}}
\newcommand {\bfE}  {{\mathbf{E}}}
\newcommand {\bfJ}  {{\mathbf{J}}}

\newcommand {\vare} {{\varepsilon}}

\newcommand {\varH} {{\mathcal{H}}}
\newcommand {\varI} {{\mathcal{I}}}
\newcommand {\varN} {{\mathcal{N}}}
\newcommand {\varO} {{\mathcal{O}}}
\newcommand {\varS} {{\mathcal{S}}}
\newcommand {\varW} {{\mathcal{W}}}

\newcommand {\bfmu}    {{\boldsymbol{\mu}}}
\newcommand {\bfgamma}   {{\boldsymbol{\gamma}}}
\newcommand {\bfsigma}   {{\boldsymbol{\sigma}}}
\newcommand {\bfeps}   {{\boldsymbol{\epsilon}}}
\newcommand {\bfnabla} {{\boldsymbol{\nabla}}}

\newcommand {\tilV}  {{\widetilde{V}}}

\newcommand {\go} { \rightarrow }

\newcommand {\Tr} {\mathop{Tr}}

\newcommand {\Ode} [2][]{ \frac{d^{#1}}{d {#2}^{#1}} }

\newcommand {\ket} [1]    { \left|#1\right\rangle }
\newcommand {\bra} [1]    { \left\langle#1\right| }
\newcommand {\braket} [2] { \left\langle\left.#1\right|#2\right\rangle }

\newcommand {\set} [1] { \left\{#1\right\} }
\newcommand {\avg}[1] { \left\langle #1\right\rangle }
\newcommand {\rcp}[1] { \frac{1}{#1} }

\newcommand {\comm} [2] { \left[#1,#2\right] }
\newcommand {\acom} [2] { \left\{#1,#2\right\} }

\begin{document}
\title{Correspondences and Quantum Description \\
  of Aharonov-Bohm and Aharonov-Casher Effects}
\author{Minchul Lee and M.Y. Choi}
\affiliation{Department of Physics, Seoul National University, Seoul 151-747, Korea}

\begin{abstract}
  We establish systematic consolidation of the Aharonov-Bohm and Aharonov-Casher effects
  including their scalar counterparts.  Their formal correspondences in acquiring
  topological phases are revealed on the basis of the gauge symmetry in non-simply
  connected spaces and the adiabatic condition for the state of magnetic dipoles.  In
  addition, investigation of basic two-body interactions between an electric charge and a
  magnetic dipole clarifies their appropriate relative motions and discloses physical
  interrelations between the effects.  Based on the two-body interaction, we also
  construct an exact microscopic description of the Aharonov-Bohm effect, where all the
  elements are treated on equal footing, i.e., magnetic dipoles are described
  quantum-mechanically and electromagnetic fields are quantized.  This microscopic
  analysis not only confirms the conventional (semiclassical) results and the topological
  nature but also allows one to explore the fluctuation effects due to the precession of
  the magnetic dipoles with the adiabatic condition relaxed.
\end{abstract}


\maketitle

\section{Introduction}

Since the celebrated work of Aharonov and Bohm \cite{Aharonov59}, topological phases in
quantum mechanical systems have been investigated intensively, both theoretically
\cite{Aharonov59,Aharonov84,SAC,Boyer87,Aharonov88,Local,Aharonov00,Aharonov02,Hagen90,He91,Goldhaber89,Santos99,Comay00,Dowling99,Comay98,DualAC}
and experimentally \cite{ABExp,ACExp,SACExp}.  There have been identified four kinds of
effects for quantum particles with electrical charges or magnetic moments, which give rise
to nontrivial and experimentally observable topological phases: The Aharonov-Bohm (AB)
effect and its scalar counterpart \cite{Aharonov59} (SAB) stand for accumulation of
topological phases of electric charges moving, without ever being in electromagnetic
fields, in vector and scalar potentials, respectively.  On the other hand, development of
topological phases by neutral particles with magnetic moments, which travel in electric
and magnetic fields though experiencing no force, is addressed as the Aharonov-Casher (AC)
\cite{Aharonov84} and the scalar AC (SAC) \cite{SAC} effects, respectively.  (In some
literature, the latter is referred to as the SAB effect.)

Since the proposal, there have been many misleading attempts to deny the existence of
topological phases in the AC and SAC effects, for example, possible classical
interpretations \cite{Boyer87,Aharonov88} and locality analysis of the AC effect
\cite{Local,Aharonov00,Aharonov02}.  Despite these, the AB and AC effects have been
pointed out similar and even equivalent \cite{Hagen90,He91} to each other.  Furthermore,
apparent difference \cite{Goldhaber89} between the AB and AC effects can be understood by
addressing the underlying principle for the generation of topological phases in both cases
\cite{He91}.  It is thus tempting to extend such topological examination to their scalar
counterparts, the SAB and SAC effects, and, further, to prove that these four complete all
kinds of topological phase acquired by a system of an electric charge and a magnetic
moment.
Such investigation gives a clue for another interrelation between the effects, viewed in
terms of the two-body interaction between an electric charge and a magnetic moment, one of
which, being congregated macroscopically, constitutes a source of electromagnetic fields.
This helps us deepen the interpretation for the SAB effect among others, which has not yet
been considered with the source included, and manifest that all the four effects simply
arise from appropriate relative motions of electric charges and magnetic dipoles.  With
the sources taken into consideration, in particular, fully quantum-mechanical
establishment of the effects can now be achieved by quantizing the electromagnetic fields;
this is to be compared with the conventional (semiclassical) setup, where a (quantum)
particle moves in {\em classical} electromagnetic fields.  Indeed, a recent study of the
AB effect, where sources for magnetic fields are treated quantum-mechanically, has yielded
results almost the same as those in the conventional setup, although the perturbative
nature of the analysis makes the conclusion short of rigorousness \cite{Santos99}.  With
rigorousness, this type of analysis would not only clarify hidden conditions for the
degree of freedom in the magnetic moment, which has invoked controversial problems in the
interpretation of the AC effect, but also make it possible to examine the possible effects
of the precession of the magnetic moment.

In this paper we establish systematic consolidation of the four topological effects.
Inspecting the condition for gauge symmetry, we find out formal correspondence between the
SAB and SAC effects, in addition to the known one between the AB and AC effects, and
demonstrate their completeness in acquiring non-trivial topological phases.  Also exposed
is the importance of the adiabatic condition for magnetic dipoles in substantiation of the
topological phase.  Further, investigation of fundamental two-body interactions between
constituent particles consisting the total system reveals physical interrelations between
the AC and SAB effects as well as between the AB and SAC effects.  Here, the relative
motions of the particles appropriate for the effects are classified.  We next consider the
quantum dynamics of the total system consisting of an electric charge and magnetic dipoles
to construct an exact quantum description of the AB effect.  By means of the path integral
representation, we analyze the charge-dipole system interacting via photons, i.e.,
quantized fields; this confirms the AB effect obtained in the semiclassical setup and
reveals the origin of the topological nature of the effect.  Finally, we justify the
adiabatic condition in real experiments and give a microscopic argument for the survival
of the nonlocal and topological phase even in the presence of weak precession of the
magnetic dipoles.

This paper is organized as follows: In \secref{sec:formal}, we examine systematically the
general condition for the formal equivalences among the effects, together with their
completeness.  In particular, the missing correspondence between the SAB and SAC effects
is revealed.  \Secref{sec:physical} is devoted to the physical interrelations between the
effects.  Presented in \secref{sec:fullquantum} is the microscopic quantum description of
the AB effect.  \Secref{sec:precession} discusses the validity of the adiabatic condition
and the effects of the precession of magnetic dipoles.  Finally, main results are
summarized in \secref{sec:conclusion}.

\section{Formal Correspondence Between Aharonov-Bohm and Aharonov-Casher
  Effects\label{sec:formal}}

The essential common property of AB and AC effects including their scalar counterparts is
acquisition of topological and nonlocal phases in force-free regions.  Such similarity
indicates structural likeness in the equations of motion for the system and the key point
for the topological phase is known to be the presence of gauge symmetry in non-simply
connected systems \cite{He91}.  The same notion can be applied to obtain the
correspondence between the SAB and SAC effects, allowing us to establish systematic
consolidation of all four kinds of the effect.  In the following we are to set up the
condition for the gauge symmetry in the relativistic formulation.

Let us first look into the AB and SAB effects.  The Dirac equation for a spin-1/2 particle
with charge $q$ and mass $m$ in external electromagnetic potential $A_\mu$ reads
\begin{equation}
  \label{eq:de_ab}
  \left(\gamma^\mu p_\mu - \frac{q}{c}\gamma^\mu A_\mu - mc\right)\psi = 0,
\end{equation}
where $\psi$ is the Dirac spinor, $p_\mu$ $(\mu=0, 1, 2, 3)$ is the covariant momentum
operator, and $\gamma^\mu$ is the Gamma matrix.  Note that under a gauge transformation
$A_\mu \go A_\mu + \partial_\mu\Gamma$, the gauge invariance principle requires the wave
function $\psi$ to transform as $\psi \go \psi' = \psi \exp\left[-(iq/\hbar c) \Gamma
\right]$ and consider a space-time region where both the electric and magnetic fields
vanish but the potential does not.  In this region, from the relations $\bfE =
-(1/c)(\partial\bfA/\partial t) - \bfnabla A_0 = 0$ and $\bfB = \bfnabla\times\bfA = 0$,
one can choose a gauge $\Gamma$ such that $A_\mu = -\partial_\mu\Gamma$, namely, $\bfA =
\bfnabla\Gamma$ and $A_0 = -(1/c)(\partial\Gamma/\partial t)$.  This indicates that in the
field-free region $(\bfE= \bfB = 0)$ $A_\mu$ becomes a pure gauge and can be gauged away.
The Dirac equation then transforms to one for a free particle:
\begin{equation}
  \label{eq:de_free}
  \left(\gamma^\mu p_\mu - mc\right)\psi' = 0,
\end{equation}
where the transformed wave function is given by
\begin{equation}
  \label{eq:psi}
  \psi' = \psi\exp\left[-\frac{iq}{\hbar c} \Gamma \right]
  = \psi \exp\left[\frac{iq}{\hbar c} \int A_\mu dx^\mu\right].
\end{equation}

\begin{figure}[tbp]
  \begin{minipage}{4cm}
    \centering
    \epsfig{file=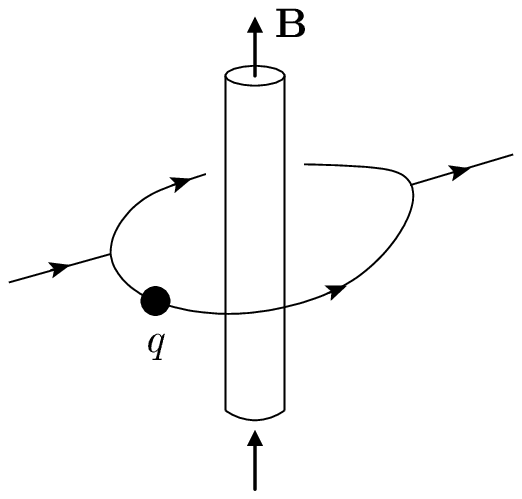,width=\textwidth}
    (a)
  \end{minipage}
  \begin{minipage}{4cm}
    \centering
    \epsfig{file=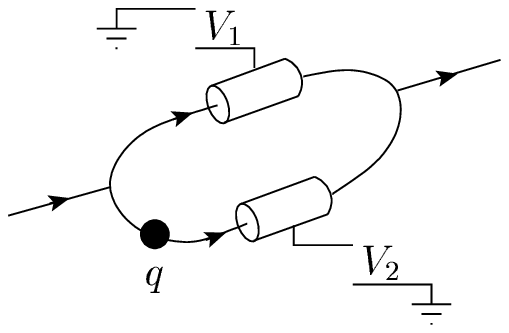,width=\textwidth}
    (b)
  \end{minipage}
  \caption{Schematic setups for (a) the AB and (b) the SAB effects.}
  \label{fig:ab}
\end{figure}

Suppose that the space-time structure for the field-free region is non-simply-connected
and has in it an impenetrable region in which the field $\bfE$ or $\bfB$ persists.  If an
electric charge circulates around a closed path in the field-free region which encloses
the impervious region, the wave function acquires a nontrivial and gauge-invariant
topological phase
\begin{equation}
  \label{eq:phase}
  \phi_{TP} \equiv \frac{q}{\hbar c} \oint A_\mu dx^\mu
  = \frac{q}{\hbar c}\oint \left(A_0cdt - \bfA\cdot d\bfx\right).
\end{equation}
In the AB setup [see \figref{fig:ab}(a)] with a time-independent magnetic field confined
inside a shielded region, the acquired phase is given by $\phi_{TP} = - (q/\hbar c)
\oint\bfA\cdot d\bfx = - (q/\hbar c) \int_S \bfB\cdot d\bfs = - (q\Phi/\hbar c)$, where
$\Phi$ is the magnetic flux passing through the area $S$ enclosed by the path.  On the
other hand, in the SAB setup [see \figref{fig:ab}(b)], the charged particle moves in a
time-dependent potential without ever being in the electric field, and one obtains the
phase $\phi_{TP} = (q/\hbar c)\oint A_0 cdt = (q/\hbar) \int \delta V(t) dt$, where
$\delta V(t)$, the potential difference at time $t$, arise from the non-zero electric
field across the prohibited space-time region.

It is now evident that the topological phase has its root in the existence of a
multiply-connected region where the potential can be gauged away so that all the effects
of the potential can be absorbed into the phase of the wave function.  In other words, in
the case of AB and SAB effects, the system must possess a topologically nontrivial region
with $\bfB = \bfE = 0$.

\begin{table*}[!t]
  \centering
  \begin{tabular}{c|c|c|c|c} \hline\hline
    & \ multiply-connected region \ & \ impervious region \ & \ dimension \
    & \ topological phase
    \\ \hline
    AB & $\bfE = \bfB = 0$ & $\bfB \ne 0$ & three & $q\Phi/\hbar c$ \\
    SAB & $\bfE = \bfB = 0$ & $\bfE \ne 0$ & three & $(q/\hbar) \int \delta V(t) dt$
    \\
    AC & $\rho = \bfJ = 0$ & $\rho \ne 0$ & two & $4\pi\mu\lambda/\hbar c$ \\
    SAC & $\rho = \bfJ = 0$ & $\bfJ \ne 0$ & one & $(\mu/\hbar)\int \delta B(t) dt$ \\
    \hline\hline
  \end{tabular}
  \caption{Properties of the multiply-connected and the impervious regions appropriate for
    the four kinds of the topological effect, together with the characteristic dimension of
    the space holding paths of the moving particle and the corresponding topological
    phase.}
  \label{tab:abc}
\end{table*}

Next the conditions for the AC and SAC effects are examined in a similar manner to the AB
effect.  We begin with the Dirac equation for a neutral spin-1/2 particle with the
magnetic dipole moment $\mu$ in the electromagnetic field $F^{\mu\nu}$:
\begin{equation}
  \label{eq:de_ac}
  \left(\gamma^\mu p_\mu  - \frac{\mu}{2c} \sigma^{\mu\nu} F_{\mu\nu} - mc\right) \psi = 0
\end{equation}
with $\sigma^{\mu\nu} \equiv \frac{i}{2} \comm{\gamma^\mu}{\gamma^\nu}$, which can be
written in the form
\begin{equation}
  \label{eq:de_ac2}
  \left( \gamma^\mu p_\mu + \frac{i\mu}{c} \bfgamma\cdot\bfE \gamma^0
    + \frac{\mu}{c} \bfsigma\cdot\bfB - mc \right) \psi = 0
\end{equation}
with the Pauli matrix $\bfsigma$.  In contrast with \eqnref{eq:de_ab} for the AB effect,
the field itself enters into the equation of motion and cannot be simply gauged away,
which implies that some restriction on the field should be imposed to make it behave like
a pure gauge.  The appropriate restriction may be uncovered by probing the condition that
the equation of motion in \eqnref{eq:de_ac} can be transformed to that for a free particle
in a multiply-connected region, given by \eqnsref{eq:de_free} and (\ref{eq:psi}) with pure
gauge potential $A_\mu$.  In general, one can write the potential in the form $A_\mu
\equiv \tau\gamma^0a_\mu$, where $a_\mu$ is a linear function of $\bfE$ and $\bfB$ and
$\tau$ is a $4\times4$ matrix to be determined in the following.  Comparison of
\eqnsref{eq:de_free} and (\ref{eq:psi}) with \eqnref{eq:de_ac2} yields the relations:
\begin{equation}
  \label{eq:comm}
  \comm{\tau}{\gamma^0} = 0 ~~\mbox{and} ~~
  \acom{\tau}{\gamma^k} = 0~\text{if $p_k\psi\ne0$} ~ (k=1,2,3)
\end{equation}
and
\begin{equation}
  i\mu \bfgamma\cdot\bfE = q\bfgamma\cdot\bfa\tau ~~\mbox{and}~~
  \mu\bfsigma\cdot\bfB = -q\gamma^0 \tau\gamma^0 a_0.
\end{equation}
The matrix $\tau$ can be expressed as a linear combination of the complete set of
$4\times4$ matrices: 1, $\gamma^\mu$, $\sigma^{\mu\nu}$, $\gamma_5 \,(\equiv
\gamma^0\gamma^1\gamma^2\gamma^3)$, and $\gamma^\mu\gamma_5$.  The commutation relations
for matrix $\tau$, given by \eqnref{eq:comm}, can be satisfied only provided at least one
of $p_k\psi$ vanishes, which implies that the dimension for the AC effect is less than
three.  In case that only one of $p_k\phi$ is zero, say, $p_3\psi = 0$, one obtains a
two-dimensional solution $\tau = \sigma^{12}$.  This leads to the potential
\begin{equation}
  \label{eq:acA}
  q\bfA = -\mu\sigma_3 (E_2,-E_1,0)
  \quad \text{and} \quad
  qA_0 = -\mu\sigma_3 B_3
\end{equation}
with $E_3 = B_1 = B_2 = 0$.  Requiring $A_\mu$ to behave like a pure gauge potential, we
have $\bfnabla\times\bfA = 0$ and $-(1/c)(\partial\bfA/\partial t) - \bfnabla A_0 = 0$,
which are followed by the charge-free condition $4\pi\rho = \bfnabla\cdot\bfE = 0$ and the
current-free condition $(4\pi/c)\bfJ = \bfnabla\times\bfB - (1/c)\partial\bfE/\partial t =
0$.  On the other hand, when two of $p_k\psi$ are set equal to zero, e.g., only $p_3\psi
\ne 0$, we obtain a one-dimensional solution of the form $\tau = i\gamma^3 \gamma_5$,
which again leads to \eqnref{eq:acA} with minor sign flips and the same charge- and
current-free conditions.

Hence, in order to develop a topological phase, a neutral spin-1/2 particle should have
its path residing on a plane (for the AC effect) or on a straight line (for the SAC
effect) in the multiply-connected region with neither charge nor current ($\rho =\bfJ =
0$) and enclosing unpenetrable regions with $\rho\ne0$ or $\bfJ\ne0$.
\begin{figure}[!t]
  \begin{minipage}{4cm}
    \centering
    \epsfig{file=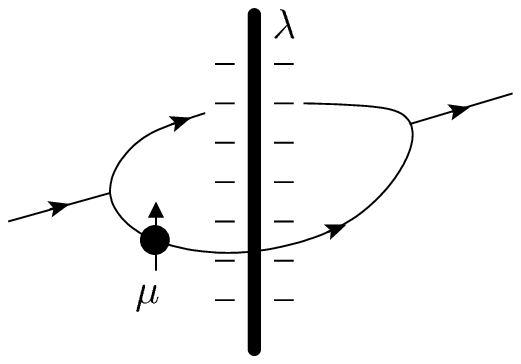,width=\textwidth}
    (a)
  \end{minipage}
  \begin{minipage}{4cm}
    \centering
    \epsfig{file=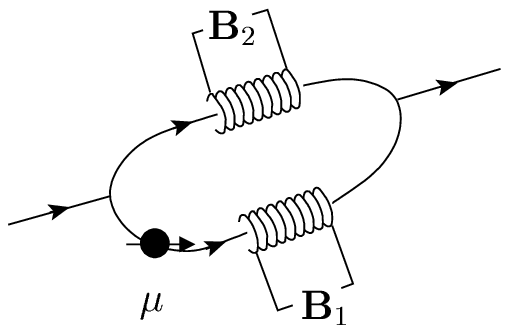,width=\textwidth}
    (b)
  \end{minipage}
  \caption{Schematic setups for (a) the AC and (b) the SAC effects.}
  \label{fig:ac}
\end{figure}
As in the AB effect, the nontrivial topological phase can be computed from
\eqnref{eq:phase} with $A_\mu$ given by \eqnref{eq:acA}.  In the AC setup [see
\figref{fig:ac}(a)], where the magnetic dipole circulates on a plane around an infinitely
long wire with charge line density $\lambda$, the topological phase is given by $\phi_{TP}
= 4\pi\mu\lambda/\hbar c$.  On the other hand, in the SAC setup shown in
\figref{fig:ac}(b), the magnetic dipole moves on a straight line under a time-varying
magnetic field aligned along the direction of the dipole, and the phase is given by
$\phi_{TP} = (\mu/\hbar)\int \delta B(t) dt$, where the magnetic-field difference $\delta
B(t)$ arises from the non-zero current flowing inside the closed path.

Tabulated in \tblref{tab:abc} are the conditions for the four kinds of topological effect
as well as the corresponding topological phases.  There the distinction between the AB and
the AC effects is made clear: In the AB and SAB effects the conditions are described in
terms of fields whereas in the AC and SAC effects the sources such as charges and currents
take the role of fields.  In addition, one can infer from the completeness of the
electromagnetic conditions listed in \tblref{tab:abc} that these four kinds encompass all
the possible nontrivial interference phenomena for a system consisting of electric charges
and magnetic dipoles.

It is straightforward to prove, under the conditions in \tblref{tab:abc} and from
\eqnref{eq:acA} with $A_\mu$ interpreted as the potential in the AB setup, that the Dirac
equations in \eqnsref{eq:de_ab} and (\ref{eq:de_ac}) are exactly equal to each other,
although in the case of the AC and SAC effects the form of the pseudo-potential $A_\mu$ is
more severely constrained.  This establishes exactly the formal correspondence between the
SAB and SAC effects as well as that between the AB and AC effects proposed already
\cite{Hagen90}, and manifests the key role of the gauge symmetry.  Note, however, that the
equivalence is possible only when one can replace the matrix $\sigma_3$ in \eqnref{eq:acA}
by one of its eigenvalues, $+1$ or $-1$.  In fact the topological phase for the AC effect
obtains the known form in \tblref{tab:abc} only under this assumption.  Since $\sigma_3$
is an operator in the spin space, such substitution implies that the spin of the magnetic
dipole should remain in one of the eigenstates or more likely in the ground state during
its motion along the path.  Namely, the magnetic dipole moves so slowly that its state
does not change, irrespectively of its interaction with the field, which is the
\textit{adiabatic condition}.  It is of interest that the adiabatic condition is also
hidden in the AB and SAB effects: The magnetic or electric field necessary for the AB or
SAB setup can be generated by a collection of stationary or moving magnetic dipoles, the
details of which are discussed in \secref{sec:physical}.  In order to build up
electromagnetic fields appropriate for the AB and SAB effects, the magnetic dipoles should
be in their ground states at all times.  Both the AB and the AC effects, therefore, are
based on adiabatic processes, and it is concluded that the adiabatic condition as well as
the gauge symmetry is essential for the topological phase.

We close this section with a comment about the force-free condition. In the AB and SAB
setups the charge moves in the field-free region so that it does not feel any force.  The
force-free condition for the AC and SAC effects, where nonvanishing fields are present,
has been proven classically in the spirit of the total momentum conservation
\cite{Aharonov88,Goldhaber89}.  Here we present a short sketch of the proof in the quantum
version.  We begin with the non-relativistic Hamiltonian with $\bfB = 0$:
\begin{equation}
  \varH = \rcp{2m} \left(\bfp-\frac{\mu}{c}\bfsigma\times\bfE\right)^2
  - \frac{\mu^2}{2mc^2} \bfE^2
\end{equation}
with the charge- and current-free conditions $\bfnabla\cdot\bfE = \bfJ = 0$.  The
Heisenberg equation of motion, with tedious applications of commutation relations, then
yields
\begin{equation}
  m \Ode[2]{t}\avg{\bfx}
  = \frac{\mu}{mc} \avg{(\bfsigma\cdot\bfnabla)(\bfE\times\bfp)}
  + \frac{\mu^2}{m\hbar c^2} \avg{\comm{\bfp\times\bfE}{\bfsigma\cdot\bfE}}.
\end{equation}
Note that with $E_z = p_z = 0$, the right-hand side has only the irrelevant $z$-component,
which vanishes if the spin is in its ground state and $\bfE$ does not depend on $z$.  We
thus observe that the force-free condition in both the AB and the AC effects comes
naturally from the gauge symmetry and the adiabatic condition.

\section{Physical Interrelation between Aharonov-Bohm and Aharonov-Casher
  Effects\label{sec:physical}}

In \secref{sec:formal} we have examined the formal equivalence between the AB and AC
effects.  The \textit{formal} equivalence stands for the agreement of the Dirac equation
describing the effects via the one-to-one correspondence like \eqnref{eq:acA}, without
regard to the physical origin.  Comparison between electromagnetic conditions listed in
\tblref{tab:abc}, however, gives us a clue for \textit{physical} interrelations between
the AB and AC effects, specifically the AB versus SAC effects and the AC versus SAB
effects.  To investigate such agreement, we consider a stationary or moving assembly of
magnetic dipoles and electric charges as sources of electromagnetic fields for the AB and
AC effects, respectively.  Namely, stationary magnetic dipoles and electric charges
generate the magnetic and electric fields, respectively, while moving ones induce the
other fields.  For the appropriate AB and AC setups, the assemblies are required to have
specific structures and motions: The AB and SAB experiments consist of a single electric
charge interacting with a structured assembly of magnetic dipoles.  In the AC and SAC
experiments the roles of the electric charge and the magnetic dipole are reversed, the
details of which is explained below.  Here the linearity of electrodynamics allows one to
write the overall interaction as a sum of two-body interactions between an electric charge
and a magnetic dipole.  Accordingly, both effects are intrinsically based on the
interaction between an electric charge and a neutral particle with magnetic moment; the
same nature of the underlying interaction suggests that the effects are physically
interrelated and equivalent to each other.  Indeed we devise a setup for the SAB effect
and newly find out the physical interrelation between the AC and SAB effects.  In the
following analysis the subscripts $q$ and $m$ are used to denote quantities pertaining to
the charge and the magnetic dipole, respectively.

We first consider the AB and SAC effects, the interrelation between which has been
demonstrated by Comay \cite{Comay00}.  For completeness here we review briefly the
argument.  In the AB experiment, the region with nonzero magnetic field can be constructed
by an infinitely long wire consisting of identical atoms or molecules with magnetic moment
$\mu$ aligned along the wire.  On the other hand, time-varying magnetic fields needed for
the SAC effect can be associated with the motion of charges along the wire of a solenoid,
which the magnetic dipole penetrates.  From these configurations, we can identify the
similar structure of the relative motion between the electric charge and the magnetic
dipole, i.e., an electric charge circling around a (stationary) magnetic dipole.  What is
different between the two effects is that in the SAC setup the motion of the charge is
confined on a plane and the magnetic dipole has another freedom of motion in the direction
perpendicular to the plane.  This freedom of motion for the magnetic dipole generates an
electric field that acts on the charges in the solenoid, but the field does not affect the
current in the solenoid; this may be easily proven by showing the line integral of the
electric field along the path of the charge vanishes as long as the magnetic dipole points
along its moving direction \cite{Comay00}, so the motion of the dipole can be ignored.
Hence the only relevant interaction in the SAC experiment is $\bfmu\cdot\bfB_q$, where
$\bfB_q$ is the magnetic field induced by the moving charge, and in the AB experiment,
$(q/c)\bfv_q\cdot\bfA_m$, where $\bfv_q$ and $\bfA_m$ are the velocity of the charge and
the vector potential generated by the magnetic dipole, respectively.  We now prove that
these two interactions are equal to each other in the origin.  Denoting
$\bfr\equiv\bfr_q-\bfr_m$, where $\bfr_q$ and $\bfr_m$ are the positions of the charge and
the magnetic dipole, respectively, we write the magnetic field induced by a moving charge
and the vector potential of a magnetic dipole:
\begin{equation}
  \bfB_q = \frac{\bfv_q}{c}\times\frac{q}{4\pi}\frac{(-\bfr)}{r^3}
  \quad\text{and}\quad
  \bfA_m = \rcp{4\pi} \frac{\bfmu\times\bfr}{r^3},
\end{equation}
which are related via
\begin{equation}
  \label{eq:ab_sac}
  \bfmu\cdot\bfB_q
  = \frac{q}{c}\bfv_q\cdot\left(\rcp{4\pi} \frac{\bfmu\times\bfr}{r^3}\right)
  = \frac{q}{c}\bfv_q\cdot\bfA_m.
\end{equation}
Hence the two-body interactions in these experiments are intrinsically the same and the
Schr\"odinger equations with the same interaction term govern the effects.  The only the
difference between them is simply whose phase change is to be measured: the electric
charge in the AB effect and the magnetic dipole in the SAC effect.

Similar relation can be found between the AC and SAB effects.  In the AC experiment, as
illustrated in \figref{fig:ac}(a), the singular structure is constructed by an infinitely
long straight wire of charges.  On the other hand, the region of a time-dependent electric
field in the SAB setup may be built up by an infinitely long hollow cylinder (i.e.,
cylindrical shell) of a finite thickness, filled with identical magnetic dipoles aligned
along the axis of the cylinder.  Rotating the cylinder along its axis generates an
electric field in the radial direction only inside the shell.  This setup can be
understood by modeling the magnetic moment as a current loop.  Summing all the nearby
current loops, one can find that azimuthal currents flow in the inner and outer surfaces
of the cylindrical shell in the opposite directions.  Although in the frame of the
cylinder charges distribute in such a way that no electric field exists, rotation of the
cylinder makes difference between the charge densities on the two surfaces in the lab
frame, which induces radial electric fields inside the shell, producing the potential
difference between the two surfaces.
In these two setups for the AC and SAB effects, another type of relative motion between an
electric charge and a magnetic dipole is identified: a magnetic dipole circling around a
(stationary) charge.  Similarly to the case of the SAC effect, there exists in the SAB
effect the freedom of motion for the charged particle along the axis of the cylinder,
which induces a magnetic field acting on the magnetic dipoles.  Being in the azimuthal
direction at all places, however, the magnetic field induced by the moving charge is
perpendicular to the magnetic dipole, having no effect on the state of the dipole at all.
Accordingly, the motion of the charge is irrelevant to our investigation, leaving the
relevant two-body interactions $-(1/c)\bfv_m\cdot\bfmu\times\bfE_q$ in the AC experiment,
where $\bfv_m$ and $\bfE_q$ are the velocity of the dipole and the electric field due to
the charge, respectively, and $qA_{0m}$ in the SAB experiment, where $A_{0m}$ is the
electric (scalar) potential induced by the moving magnetic dipole.  To examine the
relation between these two, we first note that the potential $A_{0m}$, given by the
integration of the electric field $\bfE_m$ induced by the magnetic dipole, is thus related
with the vector potential $\bfA_m$ generated by the magnetic dipole:
\begin{equation}
  \label{eq:A0}
  A_{0m} = -\int\bfE_m\cdot d\bfl
  = \int\frac{\bfv_m}{c}\times(\bfnabla\times\bfA_m)\cdot d\bfl
  = \frac{\bfv_m}{c}\cdot\bfA_m .
\end{equation}
On the other hand, the standard expressions of the electric field due to charge $q$ and
the vector potential by magnetic dipole $\bfmu$ lead to the relation
\begin{equation}
  -\frac{\bfv_m}{c}\cdot\bfmu\times\bfE_q
  = \frac{q}{c}\bfv_m\cdot\frac{\bfmu\times\bfr}{4\pi r^3}
  = \frac{q}{c}\bfv_m\cdot\bfA_m,
\end{equation}
which, together with \eqnref{eq:A0}, shows
\begin{equation}
  \label{eq:ac_sab}
  qA_{0m} = -\frac{\bfv_m}{c}\cdot\bfmu\times\bfE_q .
\end{equation}
It is thus concluded that identical two-body interactions are responsible for both the AC
and SAB effects and that these two effects are physically equivalent to each other.

\Eqnref{eq:ab_sac} indicates that the AB and SAC effects are identical in their physical
origin whereas \eqnref{eq:ac_sab} discloses similar equivalence between the AC and SAB
effects.  The difference between the corresponding effects is simply whose phase acquired
topologically is to be observed between the electric charge and the magnetic dipole.  At
this stage one may recall the proof that, based on the Galilean invariance, only the
relative motion of the charged particle and the magnetic dipole enters into the Lagrangian
of the total system and consequently, the AB and AC effects are the same in underlying
physics \cite{Aharonov84}.  There has also been an attempt to interrelate the AB/AC and
SAB/SAC effects by changing the Lorentz frame from stationary one to co-moving one
\cite{Dowling99}.  However, attention should be paid to the lack of consistent quantum
treatment in those conventional analyses, where a quantum particle moves in a classical
field, namely, the source generating the field is treated classically and immune to
quantum fluctuations.  For example, in the AB effect the electric charge is a quantum
object while the magnetic dipoles as well as its field is regarded as classical one, and
vice versa in the AC effect.  Thus in the conventional setup the difference between the AB
and AC effect is not merely the reference frame, making it desirable to give a full
quantum description.

Before closing this section, we comment about the quantum state of the magnetic dipole.
One can see that throughout the above analysis the magnetic dipole is pinned to point in a
specific direction: along the magnetic/charged wire in the AB/AC effect and along the axis
of the solenoid/cylinder in the SAB/AC effect.  This condition has also been used to argue
for the irrelevance of the extra motion of the particle in the SAB/SAC effect.  To achieve
this condition in quantum mechanics, the quantum state of the dipole must remain in an
eigenstate of the angular momentum in the supposed direction throughout the experiment.
>From this, we encounter again the adiabatic condition proposed in \secref{sec:formal},
thus the adiabatic condition is also important to establishing the interrelation between
those effects.

\section{Full Quantum Description of Aharonov-Bohm Effect\label{sec:fullquantum}}

The AB and AC effects are inherently quantum phenomena: They describe the phase acquired
by the wave function of a moving particle, either an electric charge (in the AB/SAB
effect) or a magnetic dipole (in the AC/SAC effect), which cannot be dealt with by
classical mechanics.  Accordingly, the particle is treated as a quantum object, the motion
of which is governed by the Schr\"odinger equation.  Nonetheless, the conventional theory
of the effects involves a semiclassical approximation in that the source of fields and
potentials are treated classically \cite{Santos99}.  Although the source is a macroscopic
object composed of a large number of particles (see \secref{sec:physical}), the quantum
state of each particle in the source is still important for substantiating the effect, as
discussed in \secref{sec:formal}.  Moreover, it is not possible in the conventional
semiclassical approach to study the effects of quantum fluctuations of the field and
source.  For a precise quantum interpretation, it is thus desirable to give a quantum
description of the total system including the source, and in this section we focus on the
exact quantum description of the AB effect, where magnetic dipoles as well as the moving
electric charge are treated as quantum objects and the field is quantized, expressed as a
linear combination of creation and annihilation operators of photons.  The interaction
between the charge and the magnet is then mediated via photons.  Our aim here is to
confirm the result of the conventional approach, to clarify the topological nature in the
exact quantum description, and to lay a basis for studying the precession effects of the
magnetic dipole, given in \secref{sec:precession}.

\begin{figure}
  \centerline{\epsfig{file=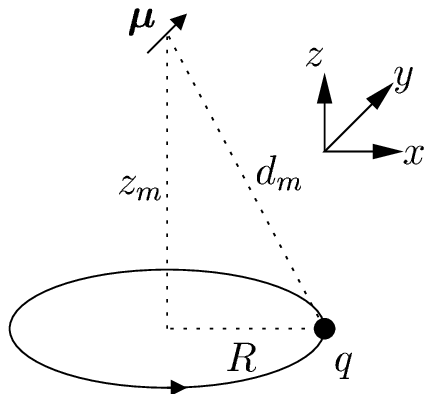,width=4.5cm}}
  \caption{A particle with charge $q$ moves around a circle on the $xy$-plane, in the
    presence of a fixed magnetic moment $\bfmu$ located on the $z$-axis, i.e., on the axis
    of the circle.}
  \label{fig:path_amagnet}
\end{figure}

As a preliminary, we first consider a system of a moving charged particle and a spatially
fixed magnetic particle.  The magnetic particle is in the eigenstate of the angular
momentum $J_u$ in the $\hat\bfu$-direction with eigenvalue $\hbar J$, so that it has a
magnetic dipole moment given by $\bfmu = g\mu_B J\hat\bfu$, where $\mu_B$ is the Bohr
magneton and $g$ the Land\`e g-factor.  In this system the particle, having charge $q$ and
mass $m$, is confined to move around a circle on the $xy$-plane whereas the magnet is
located on the $z$-axis, i.e., on the axis of the circle, as illustrated in
\figref{fig:path_amagnet}.  Thus the distance between the particle and the magnetic moment
is kept constant and given by $d_m = \sqrt{R^2+z_m^2}$, where $R$ is the radius of the
circle and $z_m$ is the distance of the magnetic dipole from the center of the circle.
The charged particle and the magnetic dipole interact with each other via virtual photon
exchange and are described by the Hamiltonian
\begin{equation}
  \label{eq:H}
  \varH
  = \rcp{2m} \left[\bfp - \frac{q}{c}\bfA(\bfr)\right]^2 - \bfmu\cdot\bfB(\bfr_m)
  + \sum_{\bfk,\hat\bfeps} \hbar\omega n_{\bfk\hat\bfeps},
\end{equation}
where $\bfp$ and $\bfr$ represent the (conjugate) momentum and position of the charge and
$\bfr_m$ the (fixed) position of the dipole.  The first term is the kinetic energy of the
charge in the presence of the vector potential $\bfA$ and the second term corresponds to
the magnetic dipole interaction at the position of the dipole.  The free photon energy is
described by the last term, where $n_{\bfk\bfeps}$ denotes the number operator for photons
with wave vector $\bfk$ and polarization $\hat\bfeps$ and the frequency is related with
the wave vector via $\omega = kc$.  In terms of photon creation/annihilation operators,
the vector potential is expressed in the Gaussian unit as
\begin{equation}
  \bfA(\bfr) = \sum_{\bfk,\hat\bfeps} \sqrt{\frac{2\pi\hbar c^2}{\omega V}} \hat\bfeps
  \left( e^{i\bfk\cdot\bfr}a_{\bfk\hat\bfeps}
  + e^{-i\bfk\cdot\bfr}a_{\bfk\hat\bfeps}^\dagger \right),
\end{equation}
where $V$ is the volume of the system.  The magnetic field is then given by $\bfB =
\bfnabla\times\bfA$.

Suppose that the particle enters into the path at time $t=t_i$ and, after circling any
number of turns, returns to the starting position at the time $t=t_f$. The state of the
particle and the dipole at the initial time is given by $\ket{\Psi;t_i} =
\ket{\psi;t_i}\ket{J}$, where $\ket{\psi;t_i}$ denotes the state of the traveling particle
and $\ket{J}$ the state of the fixed dipole.  Although any number of photons of any
momentum and polarization is allowed initially, no additional photon is assumed created or
annihilated after the travel because of its low probability.  Then the quantum
interference between the initial and the final states due to the interaction can be
obtained from the element of the time-evolution operator
\begin{eqnarray}
  \varI & \equiv & \bra{\Psi;t_i} \Tr_{\set{n_{\bfk\hat\bfeps}}}
  e^{-\frac{i}{\hbar} \int_{t_i}^{t_f} dt \varH} \ket{\Psi;t_i} \\
  \nonumber
  & = & \int d^3r_f d^3r_i
  \braket{\psi;t_i}{\bfr_f} \varW(\bfr_i,\bfr_f) \braket{\bfr_i}{\psi;t_i}
\end{eqnarray}
with
\begin{equation}
  \varW(\bfr_i,\bfr_f) \equiv \bra{J}\bra{\bfr_f}
  \Tr_{\set{n_{\bfk\bfeps}}} e^{-\frac{i}{\hbar} \int_{t_i}^{t_f} dt \varH}
  \ket{\bfr_i}\ket{J},
\end{equation}
where the trace is to be performed over all possible photon states.  For evaluating the
function $\varW$, it is natural to use the functional integral representation in terms of
the boson coherent state $\ket{\phi}$ together with the position and momentum eigenstates,
$\ket{\bfr}$ and $\ket{\bfp}$ \cite{Negele98}.  The time interval is broken into $M$
time steps of size $\vare\equiv(t_f-t_i)/M$, the closure relation
\begin{equation*}
  1 = \prod_\alpha \frac{d\phi_{\alpha,j}^*\phi_{\alpha,j}}{2\pi i}\,
  e^{-\sum_\alpha \phi_{\alpha,j}^*\phi_{\alpha,j}}
  \ket{\phi_{\alpha,j}} \bra{\phi_{\alpha,j}}
  \otimes
  d^3r_j \ket{\bfr_j}\bra{\bfr_j},
\end{equation*}
where $\alpha$ stands for the photon state $(\bfk,\bfeps)$, is inserted at the $j$th time
step $(j=1,\ldots,M{-}1)$, and the periodic boundary conditions are used: $\phi_{\alpha,0}
= \phi_\alpha$ and $\phi_{\alpha,M}^* = \phi_\alpha^*$ for $\phi$, and $\bfr_0 = \bfr_i$
and $\bfr_M = \bfr_f$ for $\bfr$.  We then obtain
\begin{equation}
  \varW = \lim_{M\go\infty}
  \int
  \prod_{j=1}^M \prod_\alpha \frac{d\phi_{\alpha,j}^*d\phi_{\alpha,j}}{2\pi i}
  \prod_{j=1}^{M{-}1}d^3r_j \prod_{j=1}^M\frac{d^3p_j}{(2\pi\hbar)^3}
  \,e^{-\varS}
\end{equation}
where the action is given by
\begin{eqnarray}
  \nonumber
  \varS & = &
  \sum_{j=1}^M
  \left\{
    \sum_\alpha
    \left[\phi_{\alpha,j}^*\phi_{\alpha,j}
      - (1-i\vare\omega)\phi_{\alpha,j}^*\phi_{\alpha,j{-}1} \right]
  \right.\\
  \label{eq:action}
  & & \qquad\quad\mbox{}
  - \frac{i\vare}{\hbar} \sum_\alpha
  \left( F_{\alpha,j}^* \phi_{\alpha,j{-}1} + F_{\alpha,j} \phi_{\alpha,j}^* \right) \\
  \nonumber
  & & \left.\quad\qquad\mbox{}
    - \frac{i}{\hbar} \bfp_j\cdot(\bfr_j {-} \bfr_{j{-}1})
    + \frac{i\vare}{\hbar}\frac{\bfp_j^2}{2m}
  \right\}.
\end{eqnarray}
with
\begin{eqnarray}
  F_{\alpha,j} & \equiv & \sqrt{\frac{2\pi\hbar c^2}{\omega V}}
  \left[\frac{q}{mc} \bfeps\cdot\bfp_j\, e^{-i\bfk\cdot\bfr_{j{-}1}} \right. \nonumber\\
  & & \qquad\quad\qquad\mbox{}  \left.
    - i (\bfmu\cdot\bfk\times\bfeps)\, e^{-i\bfk\cdot\bfr_m} \right].
\end{eqnarray}
Here we have disregarded the $\bfA^2$-term describing a two-photon process since its
contribution is usually much smaller than the single-photon process.

Since the action in \eqnref{eq:action} is quadratic in the complex variables
$\phi_{\alpha,j}$, simple Gaussian integration can be performed over $\phi_{\alpha,j}$,
yielding
\begin{equation}
  \varW = \lim_{M\go\infty}
  \int
  \prod_{j=1}^{M{-}1}d^3r_j \prod_{j=1}^M\frac{d^3p_j}{(2\pi\hbar)^3}
  \,e^{-\varS}
\end{equation}
with
\begin{equation}
  \varS =
  -\sum_\alpha \frac{i\vare}{\hbar} \frac{|\sum_{j=1}^M F_{\alpha,j}|^2}{M\hbar\omega}
  - \sum_{j=1}^M \left[ \frac{i}{\hbar} \bfp_j{\cdot}(\bfr_j {-} \bfr_{j{-}1})
    {-} \frac{i\vare}{\hbar} \frac{\bfp_j^2}{2m} \right]
\end{equation}
up to $\varO(\vare)$.  Here we have omitted a factor coming from the time evolution of
photons in which we are not interested.  The $|\sum F|^2$-term contains three
contributions: two from the self-energies of the charge and the dipole and one from the
charge-dipole interaction.  The contribution from the self-energy of the dipole is
irrelevant to our analysis and one from the self-energy of the charge is extremely small
by several orders of magnitude compared with other terms.  Disregarding these two and
summing over the photon polarization and momentum by converting $V^{-1}\sum_\bfk$ into
$(2\pi)^{-3}\int d^3k$, we write the action in the form
\begin{eqnarray}
  \label{eq:branch}
  \varS
  & = &  \sum_{j=1}^M
  \left[
    \frac{i\vare}{\hbar} \frac{\bfp_j^2}{2m}
    - \frac{i}{\hbar} \bfp_j\cdot(\bfr_j - \bfr_{j{-}1}) \right. \\
  \nonumber
  & & \qquad\qquad\left.\mbox{}
    - \frac{i\vare}{\hbar}
    \frac{q}{mc}
    \frac{\bfp_j\cdot\bfmu\times(\bfr_{j{-}1}-\bfr_m)}{|\bfr_{j{-}1}-\bfr_m|^3}
  \right].
\end{eqnarray}
Here the last term may be interpreted as the interaction between the dipole and the
magnetic field $\bfB_q$ generated by the charge in motion:
\begin{equation}
  \frac{q}{mc} \frac{\bfp_j\cdot\bfmu\times(\bfr_{j{-}1}-\bfr_m)}{|\bfr_{j{-}1}-\bfr_m|^3}
  =
  \bfmu\cdot\bfB_q(\bfr_m {-}\bfr_{j{-}1}),
\end{equation}
where the Biot-Savart law has been used to give $\bfB_q(\bfr) =(q/mc) \bfp\times\bfr/r^3$.
This confirms the interrelation proposed in \secref{sec:physical} between the AB and SAC
effects in the consistent quantum-mechanical description.

Tracing out the momentum variables, we obtain the element
\begin{equation}
  \label{eq:product}
  \varI =
  \lim_{M\go\infty}
  \rcp{\varN}
  \int \prod_{j=0}^M d^3r_j
  \braket{\psi;t_i}{\bfr_M} e^{-\varS} \braket{\bfr_0}{\psi;t_i}
\end{equation}
with
\begin{eqnarray}
  \nonumber
  \varS
  & = &
  - \frac{i}{\hbar} \sum_{j=1}^M
  \left[
    \frac{m}{2\vare}(\bfr_j{-}\bfr_{j{-}1})^2
    +
    \frac{\vare q^2}{2mc^2} \frac{[\bfmu\times(\bfr_{j{-}1}{-}\bfr_m)]^2}
    {|\bfr_{j{-}1}-\bfr_m|^6} \right. \\
  & & \left.\qquad\quad\mbox{}
    +
    \frac{q}{c}\frac{(\bfr_j-\bfr_{j{-}1})\cdot\bfmu\times(\bfr_{j{-}1}-\bfr_m)}
    {|\bfr_{j{-}1}-\bfr_m|^3}
  \right],
\end{eqnarray}
where $\varN$ is an appropriate normalization constant.
Without any geometrical assumption, the element $\varI$ in \eqnref{eq:product} contains
all the contributions coming from any path of the particle.  We now take into account only
the path shown in \figref{fig:path_amagnet}, especially with $\hat\bfu=\hat\bfz$, i.e.,
$\bfmu \equiv g\mu_B J \hat\bfz$.  In the polar coordinate $\bfr_j =
R(\cos\theta_j,\sin\theta_j,0)$, we replace the integral over $\bfr_j$ with one for
$\theta_j$, and get
\begin{eqnarray}
  \nonumber
  \varI & = &
  \lim_{M\go\infty}
  \frac{e^{i\varphi_1(z_m)}}{\varN}
  \int_0^{2\pi} \prod_{j=0}^M \frac{d\theta_j}{2\pi}
  \braket{\psi;t_i}{\theta_M}  \braket{\theta_0}{\psi;t_i} \\
  & & \qquad\qquad\qquad\qquad\quad\mbox{}
  \times e^{\sum_{j=1}^M V(\theta_j-\theta_{j{-}1})},
\end{eqnarray}
where we have defined
\begin{eqnarray}
  \varphi_1(z_m) & \equiv & \frac{(t_f-t_i)q^2\mu^2}{2mc^2 d_m^4} \nonumber \\
  V(\theta) & \equiv & i\kappa (1-\cos\theta) + \frac{i\phi_1(z_m)}{2\pi} \sin\theta
\end{eqnarray}
with $\kappa \equiv mR^2/\hbar\vare$ and
\begin{equation}
  \label{eq:phaseshift1}
  \phi_1(z_m) \equiv \frac{2\pi q\mu R^2}{\hbar c d_m^3}.
\end{equation}
Since $e^{V(\theta)}$ is periodic in $\theta$, it can be expressed as a Fourier series:
$e^{V(\theta)} = \sum_{s=-\infty}^\infty e^{is\theta} e^{\tilV(s)}$, where $e^{\tilV(s)} =
\int_{-\pi}^\pi (d\theta/2\pi) e^{-is\theta}e^{V(\theta)}$.  Using this Fourier expansion
of $e^{V(\theta_j-\theta_{j{-}1})}$ and integrating over $\theta_j$ for $1\le j\le M{-}1$,
we obtain
\begin{eqnarray}
  \nonumber
  \varI & = &
  \lim_{M\go\infty}
  \frac{e^{i\varphi_1(z_m)}}{\varN}
  \int_0^{2\pi} \frac{d\theta_0d\theta_M}{(2\pi)^2}
  \braket{\psi;t_i}{\theta_M}  \braket{\theta_0}{\psi;t_i} \\
  \label{eq:general}
  & & \qquad\mbox{}\times
  \sum_{n=-\infty}^\infty \int_{-\infty}^\infty d\xi
  e^{M\tilV(\xi)} e^{in(\theta_M-\theta_0) + 2\pi i n\xi},
\end{eqnarray}
where we have used the fact that the integral $\int_0^{2\pi} (d\theta/2\pi) e^{is\theta}$
gives rise to the Kronecker delta function $\delta_{s,0}$ and applied the Poisson
summation formula on the summation variable.

Let us consider the case where the charge enters initially into the circle at $\theta =
0$, such that the initial wave function is given by the delta function:
$\braket{\theta}{\psi;t_i} = 2\pi\delta(\theta)$.  Then $\theta_0$ and $\theta_M$ are set
equal to zero and we have, from the definition of $\tilV(\xi)$,
\begin{align}
  & \int_{-\infty}^\infty d\xi e^{M \tilV(\xi) + 2\pi i n\xi} \\
  \nonumber
  & =
  \int_{-\infty}^\infty \!\!d\xi
  \left[\int_{-\pi}^\pi \!\frac{d\theta}{2\pi} e^{-i\xi\theta}
    e^{i(\kappa(1{-}\cos\theta) + (\phi_1/2\pi)\sin\theta)}\right]^M
  \!\!\! e^{2\pi i n\xi} ,
\end{align}
where in the limit $M\go\infty$ or $\kappa\go\infty$, the integrand except near $\theta=0$
oscillates rapidly to be inconsequential so that we may approximate
$\sin\theta\approx\theta$.  Shifting the variable $\xi$ by $\phi_1(z_m)/2\pi$, we obtain
\begin{equation}
  \label{eq:finalI}
  \varI = e^{i\varphi_1(z_m)} \sum_n e^{in\phi_1(z_m)} \varI_0(n),
\end{equation}
where $n$ represents the winding number of the charge around the dipole and $\varI_0(n)$
is the interference amplitude for a free particle (with the winding number $n$) in the
absence of the magnetic field. The phase shift by the photon-mediated interaction between
the charge and the dipole thus consists of the winding-number dependent part and the
time-dependent (dynamical) one: $n\phi_1(z_m)+\varphi_1 (z_m)$.  Note that the former as
well as the latter still depends on the size and the relative position of the path, and is
not topological at all.

\begin{figure}
  \centerline{\epsfig{file=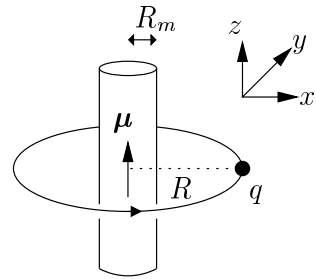,width=4.5cm}}
  \caption{A particle with charge $q$ circles around an infinitely long cylindrical
    magnet along the $z$-direction.  As before, the trajectory of the particle is confined
    on the $xy$-plane.}
  \label{fig:path_magnet}
\end{figure}

We now investigate the original AB setup shown in \figref{fig:path_magnet}. We assume that
the magnet has an infinitely long cylindrical structure with radius $R_m \,({<}R)$ and is
composed of $N$ identical atoms or molecules per unit volume, each of which has the same
dipole moment given before. Then the dipole interaction term $-\bfmu\cdot\bfB(\bfr_m)$ in
the Hamiltonian (\ref{eq:H}) is replaced by $-\sum_{\bfr_m} \bfmu\cdot\bfB(\bfr_m)$.
Taking the $z$-axis along the cylindrical axis of the magnet, namely, $\hat\bfu=\hat\bfz$,
we follow the same procedure as that for a single dipole to reach \eqnref{eq:branch} with
the additional summation over $\bfr_m$ in the third term; self-energy contribution of the
magnet is also modified but it is irrelevant as before.  Converting the summation
$\sum_{\bfr_m}$ into the integral $N\int_{\bfr_m} d^3\bfr_m$ extended to the volume of the
magnet and performing the integration, we obtain, in the place of \eqnref{eq:branch},
\begin{eqnarray}
  \nonumber
  \varS
  & = & \sum_{j=1}^M
  \left[
    \frac{i\vare}{\hbar} \frac{\bfp_j^2}{2m}
    - \frac{i}{\hbar} \bfp_j\cdot(\bfr_j - \bfr_{j{-}1}) \right. \\
  & & \qquad\qquad\left.\mbox{}
    + \frac{i\vare}{\hbar}
    \frac{q\Phi}{2\pi mc}
    \frac{\bfp_j\cdot\bfz\times\bfr_{j{-}1}}{\bfr_{j{-}1}^2}
  \right],
\end{eqnarray}
where $\Phi \equiv (4\pi N\mu) (\pi R_m^2)$ is the magnetic flux inside the magnet.
Following the remaining procedure leads to \eqnref{eq:finalI} with the replacement of
$\phi_1$ and $\varphi_1$ by
\begin{equation}
  \phi_{AB} \equiv \frac{q\Phi}{\hbar c}
  \quad\text{and}\quad
  \varphi \equiv \frac{(t_f-t_i)\hbar\phi_{AB}^2}{8\pi^2 mR^2},
\end{equation}
respectively.  Referring to the data used in an AB interferometry experiment \cite{ABExp},
we observe that the dynamical part $\varphi$ is of the order of $\lambda_{deB}/2\pi R$,
where $\lambda_{deB}$ is the de Broglie wavelength of the charge, and negligibly small by
several orders of magnitude in comparison with $\phi_{AB}$.  On the other hand, for
winding number $n$, the topological phase shift is given by $n\phi_{AB}$, which is the
same as that in the conventional AB effect.

This result can also be obtained from the phase shift generated by a single dipole in
\eqnref{eq:phaseshift1} in the following way: Consider a charged particle circling around
an infinitely long wire along the $z$-direction, with the linear density $N$ of magnetic
dipoles.  The total phase shift for winding number $n$ is then given by
\begin{equation}
  \sum_{\bfr_m} n\phi(z_m) = N \int_{-\infty}^\infty dz_m n\phi(z_m)
  = \frac{n q (4\pi \mu N)}{\hbar c}.
\end{equation}
Identifying the magnetic flux $\Phi = 4\pi\mu N$ here, we obtain the same topological
phase $n\phi_{AB}$, which manifests that the topological effect arises from the
non-simply-connected geometry of the space.

\section{Precession of Magnetic Dipoles\label{sec:precession}}

In previous sections we have assumed that magnetic dipoles are inert and retain their
ground state throughout experiment, regardless of their interactions with electromagnetic
fields or potentials.  Such inertness of magnetic dipoles is a crucial ingredient in
acquiring well-defined topological phases and interrelations between the AB and AC
effects.  Fortunately, this condition can usually be justified in real experimental
situations.  In AB/SAB experiment, the dipoles in the (cylindrical) magnet have very
strong ferromagnetic interactions between them, giving rise to negligibly small transition
probability to excited states by the interaction with the magnetic field generated by an
electric charge in motion \cite{Comay00}.  Similarly, the neutron used as a neutral
particle with nonzero magnetic moment in AC/SAC experiment hardly excites to higher-energy
states, with the energy level several hundred MeV above the ground state, by the very weak
interaction with the magnetic field of charges \cite{Comay98}.  For this reason, the
magnetic dipole may be regarded as an inert object whose state is not affected by the
field of the charge in relative motion, and the adiabatic condition is satisfied.

Then what if the proviso for the inertness is relaxed?  It has been shown that, in the
opposite limit where the magnetic dipole is replaced by a classical object such as a thin
circular pipe containing a charged fluid, the phase shift for the AB and SAC effects
disappears completely \cite{Comay98,Comay00}.  The reason is that the kinetic energy of
charges in the classical object changes in the presence of an external magnetic field in
such a way that it cancels out exactly the contribution of the magnetic field to the
action, making the action independent of the field.

More profoundly, in the absence of the adiabatic condition, one cannot replace the matrix
$\sigma_3$ in \eqnref{eq:acA} by its eigenvalue and has to treat the spin as a dynamic
quantum variable.  As shown in \secref{sec:physical}, the magnetic moment always couples
to the magnetic field induced by the charge in relative motion in both the AB/SAB and the
AC/SAC effects.  Consequently, the magnetic moment precesses around the magnetic field and
the accumulation of the quantum phase shift is accompanied by the local precession.  There
has been an argument against the nonlocality of the AC/SAC effect, based on the
commutativity and the linear relation between the phase shift and the precession angle
which can be locally observable \cite{Local}. If we follow their assertion, however, the
interrelation demonstrated in \secref{sec:physical} lead us to conclude that the AB/SAB
effect is also local.  Recently, flaws in their argument have been pointed out
\cite{Aharonov00,Aharonov02}: Since the phase shift and the precession angle are
noncanonical variables, commutativity between them does not necessarily imply mutual
observability \cite{Aharonov75} and instead the observation of the local precession
induces uncertainty in the phase shift. (Here a noncanonical variable means an observable
depending on the Hamiltonian of the system.)  Therefore even if the magnetic dipole can
precess, it should not hurt the nonlocal and topological nature of the AB and AC effects.

However, it can give a quantitative correction in the phase shift acquired since the
precession of the dipole due to the motion of the charge in turn affects the state of the
charge.  Let us consider the AB effect microscopically as in \secref{sec:fullquantum} and
now suppose that the dipole can precess or be excited energetically as it interacts via
photons.  Then the entire system will culminate in some entangled state of the charge and
the dipole just before the charge interferes with its initial state or in other words it
hits the screen in the AB interferometry.  With respect to each final state of the dipole,
the charge may acquire different phases and the states of different phases may interfere
with each other to obscure the interference pattern on the screen.  The correction in the
interference, however, may be small compared with the topological phase because it
involves at least two-photon processes, one from the charge to the dipole and one in
return; the main contribution to the phase shift comes from the one-photon process from
the dipole to the charge.  Hence we may neglect the effects of the precession on the
charge interference pattern as long as the precession of the dipole moment is not too
large, which is expected to be well satisfied in experiment.

\section{Conclusions\label{sec:conclusion}}

In this work we have established systematic unification of the AB and AC effects together
with their scalar counterparts, making clear two kinds of correspondence: formal and
physical ones.  There exist formal equivalences between the AB and AC effects and between
the SAB and SAC effects.  They are based on the multiply-connected geometry of the space,
where the potential or the field can be gauged away so that their effects can be absorbed
into the phase of the wave function, together with the adiabatic condition for the quantum
state of magnetic dipoles. The list of electromagnetic conditions for these four effects
indicates their completeness in topological phases which can be acquired by systems of
electric charges and magnetic dipoles.  In addition, by devising the structure of the
source for the SAB effect which has been missing, we have also completed the analysis of
the two-body interaction and found that the same two-body interactions between an electric
charge and a magnetic dipole are responsible for the AB and the SAC effects and for the AC
and SAB effects, which suggests the physical interrelations between them.

The importance of quantum treatment of the total system encompassing the source has been
appreciated during the consolidating analysis.  Accordingly, for precise quantum results,
we have given an exact description of the AB effect by mean of the path integral
representation.  Beginning with a system consisting of an electric charge and a magnetic
dipole, interacting with each other via virtual photons, we have confirmed the result of
the conventional approach and manifested the topological nature of the AB effect.
Finally, the adiabatic condition has been argued to hold in real experiment and the
survival of the nonlocality of the topological phase been discussed in the presence of the
precession of the magnetic dipole.

Recently, by exploiting the electromagnetic duality in the Maxwell's equations, some
authors have suggested new kinds of gedanken topological effects for a magnetic monopole
or an electric dipole \cite{Dowling99,DualAC}.  Since their equations of motion have
exactly the same structure as those of their dual effects, i.e., the AB and AC effects,
all the relations between the effects found in this paper can be directly applied to them
simply by substituting magnetic charges and electric dipoles for electric charges and
magnetic dipoles. In addition, following our argument, one can find the structure of their
scalar counterparts systematically \cite{Aharonov00}.  Hence our results encompass
essentially all the topological effects implied in the quantum electromagnetic theory.

\acknowledgments

M.Y.C. thank the Korea Institute for Advanced Study for hospitality, where part of this
work was accomplished.  We thank M.-S. Choi for useful discussions and acknowledge the
partial support by the SKOREA Program and by the BK21 Program.


\begin{thebibliography}{99}

\bibitem{Aharonov59}
  Y. Aharonov and D. Bohm, Phys. Rev. {\bf 115}, 485 (1959).

\bibitem{Aharonov84}
  Y. Aharonov and A. Casher, Phys. Rev. Lett. {\bf 53}, 319 (1984).

\bibitem{SAC}
  A. Zeilinger, in \textit{Fundamental Aspects of Quantum Theory}, edited by V. Gorini and
  A. Frigerio, NATO Advanced Study Institute Series B, Vol. 144 (Plenum, New York, 1985), p. 311;
  J. Anandan, Phys. Lett. A {\bf 138}, 347 (1989); {\bf 152}, 504 (1991).

\bibitem{Boyer87}
  T. H. Boyer, Phys. Rev. A {\bf 36}, 5083 (1987).

\bibitem{Aharonov88}
  Y. Aharonov, P. Pearle, and L. Vaidman, Phys. Rev. A {\bf 37}, 4052 (1988).

\bibitem{Local}
  M. Peshkin and H. J. Lipkin, Phys. Rev. Lett. {\bf 74}, 2847 (1995);
  P. Hyllus and E. Sj\"oqvist, {\it ibid}. {\bf 89}, 198901 (2002).

\bibitem{Aharonov00}
  Y. Aharonov and B. Reznik, Phys. Rev. Lett. {\bf 84}, 4790 (2000).

\bibitem{Aharonov02}
  Y. Aharonov and B. Reznik, Phys. Rev. Lett. {\bf 89}, 198902 (2002).

\bibitem{Hagen90}
  C. R. Hagen, Phys. Rev. Lett. {\bf 64}, 2347 (1990).

\bibitem{He91}
  X. G. He and B.H.J. McKellar, Phys. Lett. B {\bf 256}, 250 (1991).

\bibitem{Goldhaber89}
  A. S. Goldhaber, Phys. Rev. Lett. {\bf 62}, 482 (1989).

\bibitem{Santos99}
  E. Santos and I. Gonzalo, Europhys. Lett. {\bf 45}, 418 (1999).

\bibitem{Comay00}
  E. Comay, Phys. Rev. A {\bf 62}, 042102 (2000).

\bibitem{Dowling99}
  J. P. Dowling, C. P. Williams, and J. D. Franson, Phys. Rev. Lett. {\bf 83}, 2486 (1999).

\bibitem{Comay98}
  E. Comay, Phys. Lett. A {\bf 250}, 12 (1998).

\bibitem{DualAC}
  X. G. He and B.H.J. McKellar, Phys. Rev. A {\bf 47}, 3424 (1993);
  M. Wilkens, Phys. Rev. Lett. {\bf 72}, 5 (1994);
  J. Yi, G. S. Jeon, and M.Y. Choi, Phys. Rev. B {\bf 52}, 7838 (1995);
  T.-Y. Lee, Phys. Rev. A {\bf 62}, 064101 (2000).

\bibitem{ABExp}
  A. Tonomura, N. Osakabe, T. Matsuda, T. Kawasaki, J. Endo, S. Yano, and H. Yamada,
  Phys. Rev. Lett. {\bf 56}, 792 (1986), and earlier references cited therein.

\bibitem{ACExp}
  A. Cimmino, G. I. Opat, A. G. Klein, H. Kaiser, S. A. Werner, M. Arif, and R. Clothier,
  Phys. Rev. Lett. {\bf 63}, 380 (1989);
  K. Sangster, E. A. Hinds, S. M. Barnett, and E. Riis, {\it ibid}. {\bf 71}, 3641
  (1993).

\bibitem{SACExp}
  B. E. Allman, A. Cimmino, G. I. Opat, A. G. Klein, H. Kaiser, and S. A. Werner,
  Phys. Rev. Lett {\bf 68}, 2409 (1992);
  B. E. Allman, W.-T. Lee, O. I. Motrunich, and S. A. Werner, Phys. Rev. A {\bf 60}, 4272
  (1999).

\bibitem{Negele98}
  See, e.g., J. W. Negele and H. Orland, \textit{Quantum Many-Particle Systems}
  (Addison-Wesley, Redwood City, 1998).

\bibitem{Aharonov75}
  Y. Aharonov and J. L. Safko, Ann. Phys. (N.Y.) {\bf 91}, 279 (1975).

\end{thebibliography}
\end{document}